\documentclass[%
 reprint,
superscriptaddress,
 amsmath,
 amssymb,
 aps,
]{revtex4-2}
\usepackage{pdfpages}
\makeatletter
\AtBeginDocument{\let\LS@rot\@undefined}
\makeatother

\usepackage{graphicx}
\usepackage{comment}
\usepackage{dcolumn}
\usepackage{bm}
\usepackage{hyperref}

\usepackage{xfrac}
\usepackage{siunitx}
\usepackage[T1]{fontenc} 
\begin{document}
\preprint{APS/123-QED}

\title{High fidelity optical readout of a nuclear spin qubit in Silicon Carbide}

\author{Erik Hesselmeier}
\thanks{These authors contributed equally}
 \affiliation{3rd Institute of Physics, IQST, and Research Center SCoPE, University of Stuttgart, Stuttgart, Germany}
\author{Oliver von Berg}
\thanks{These authors contributed equally}
 \affiliation{3rd Institute of Physics, IQST, and Research Center SCoPE, University of Stuttgart, Stuttgart, Germany}

\author{Pierre Kuna}
\thanks{These authors contributed equally}
 \affiliation{3rd Institute of Physics, IQST, and Research Center SCoPE, University of Stuttgart, Stuttgart, Germany}
 
  
\author{Wolfgang Knolle}
\affiliation{Department of Sensoric Surfaces and Functional Interfaces, Leibniz-Institute of Surface Engineering (IOM), Leipzig, Germany}

\author{Florian Kaiser}
  \affiliation{Materials Research and Technology (MRT) Department, Luxembourg Institute of Science and Technology (LIST), 4422 Belvaux, Luxembourg}
  \affiliation{University of Luxembourg, 41 rue du Brill, L-4422 Belvaux, Luxembourg}

\author{Nguyen Tien Son}
 \author{Misagh Ghezellou}
\author{Jawad Ul-Hassan}
\affiliation{Department of Physics, Chemistry and Biology, Linköping
 	University, Linköping, Sweden}

\author{Vadim Vorobyov}
\email[]{v.vorobyov@pi3.uni-stuttgart.de}
 \affiliation{3rd Institute of Physics, IQST, and Research Center SCoPE, University of Stuttgart, Stuttgart, Germany}  

\author{J\"org Wrachtrup}
 \affiliation{3rd Institute of Physics, IQST, and Research Center SCoPE, University of Stuttgart, Stuttgart, Germany}
 \affiliation{Max Planck Institute for solid state physics, Stuttgart, Germany}


\begin{abstract}
Quantum state readout is a key requirement for a successful qubit platform. 
In this work we demonstrate a high fidelity quantum state readout of a V2 center nuclear spin based on a repetitive readout technique.
We demonstrate up to 99.5$\,\%$ readout fidelity and 99$\,\%$ for state preparation. 
Using this efficient readout we initialise the nuclear spin by measurement and demonstrate its Rabi and Ramsey nutation. 
Finally, we use the nuclear spin as a long lived memory for quantum sensing application of weakly coupled diatomic nuclear spin bath.  
\end{abstract}

\maketitle

\section{Introduction}
Efficient readout of quantum states is at the heart of modern quantum technology. It allows to perform single shot measurements, i.e. the readout of the eigenstate of an individual quantum system reaching nearly unity fidelities \cite{myerson2008high, pla2013high,pla2013high}. 
Among numerous quantum technological platforms \cite{devoret2013superconducting, saffman2010quantum,wineland2011quantum}, optically active spin-qubits \cite{Awschalom_2018}  are one of the emerging solid state systems suitable for realisation of optically interfaced spin registers \cite{kalb2017entanglement}.
Among them, the Silicon Vacancy (V2) center in 4H-SiC is an attractive solid state platform combining both, excellent optical and host material properties. 
It possesses Fourier-limited optical transitions resilient to phonon broadening up to $T=\SI{20}{K}$ \cite{Udvarhelyi2020} and has tolerable optical coherence properties in nanophotonic structures \cite{babin2022fabrication, Lukin:2023aa} making it suitable for establishing an efficient spin-photon interface. 
Yet, low optical cyclicity and long lived meta-stable states \cite{liu2023silicon} have made a single shot readout of a single electron spin state impossible so far.

Achieving an efficient spin readout is a major milestone advancing various applications. 
High fidelity optical readout of spin qubit states is important for quantum information processing, e.g. for feedback steps in error correction \cite{waldherr2014quantum, cramer2016}. 
Additionally, a universal enhancement of the electron spin readout \cite{steiner2010universal}, as achieved here, boosts the sensitivity and speeds up the exploration of a rich nuclear spin bath for Hamiltonian estimation of multispin registers \cite{abobeih2019atomic}. 
An efficient spin readout also increases the success probability of high fidelity heralded spin-photon entanglement thus increasing its distribution rate. 
In the past, mapping of a short lived electron spin state to a longer lived one such as charge state \cite{elzerman2004single,hanson2006single, shields2015efficient, anderson2022five} or nuclear spin states \cite{neumann2010single, dreau2013single} were used to increase the signal to noise ratio beyond unity, required for single shot readout.
Additionally, an increase of the collection efficiency and emission rate of the emitted photons via photonic integration and the Purcell enhancement \cite{lukin2020integrated} has been used to improve the readout fidelity. 

\begin{figure}
	\centering
	\includegraphics[width=\columnwidth]{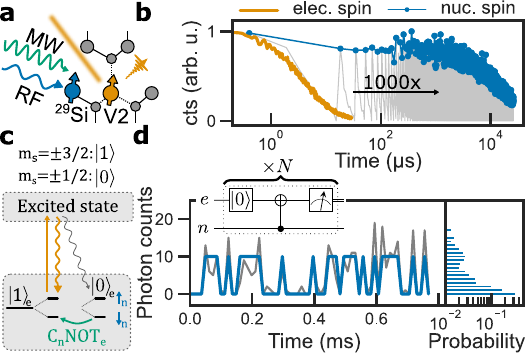}
	\caption{\textbf{(a)}  V2 center in 4H-SiC with nearby nuclear spin of $^{29}$Si \textbf{(b)} Photon emission under resonant excitation. Electron spin (blue) decays within a few microseconds while the nuclear spin state is preserved for over a millisecond. \textbf{(c)} Hyperfine energy level diagram \textbf{(d)} Real-time trace of nuclear spin quantum jumps}
	\label{fig1}
\end{figure}

In this work we perform an efficient readout of an ancillary neighbouring $^{29}$Si nuclear spin ($I=1/2$) via repetitive optical readout and mapping onto the electron spin state. 
We analyze the stochastic switching dynamics of the nuclear spin under the readout and extract its main properties and use them to optimise the readout parameters.
We find that state preparation fidelity of $\mathcal{F}_{init}$=99\% is achievable and a readout fidelity of $\mathcal{F}_{read}$=92\% in case of no data loss.
With postselection and data loss one could further improve the fidelity to $\mathcal{F}_{read}$=99.5\% with $\eta=10\%$ success rate. 
\begin{figure*}
	\centering
	\includegraphics[width=\textwidth]{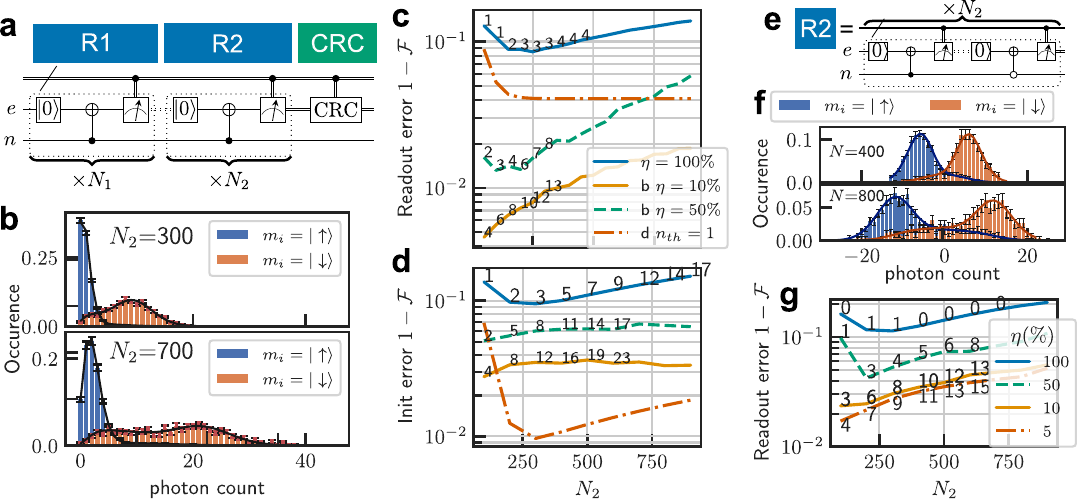}
		\caption{\textbf{(a)} Two point measurement scheme consisting of two readouts R1 and R2 and a charge resonant check (CRC) \textbf{(b)} Photon count histogram for the second readout R2 conditioned on charge-resonance check and first readout R1 \textbf{(c)} Infidelity of the readout using the scheme described in (a). Numbers label the threshold photon count \textbf{(d)} Infidelity of state preparation by readout presented in (a). \textbf{(e)} Scheme for alternating addressing of both nuclear spin states. \textbf{(f)} Distribution of difference of photon counts in alternating addressing of spin states. \textbf{(g)} Infidelity of readout of the scheme in (e).}
		\label{fig2}
\end{figure*}

Furthermore, our model gives insights on the decay mechanisms of the memory. 
In particular, we see that the bright state decays tenfold faster, while the dark state is more stable under readout, which we use in the end for state preparation. 
Finally, we demonstrate the application of this readout for probing the coherence properties of the nuclear spin memory and quantum sensing of weakly coupled nuclear spins. 
Our results mark a key achievement for the V2 system bringing it closer to a fully fledged qubit platform. 

\section{Results}
For optical addressing we use a single vacancy of silicon (V2 center) in the 4H-SiC crystal depicted in Fig. \ref{fig1}a.

Its electron spin state can be probed by resonant optical excitation resulting in $\sim 0.02$ photons per readout. This low photon rae and the fact that optical excitation resets the spin into the dark state prevents an efficient electron state readout (Fig. \ref{fig1}b.)
Neighbouring nuclear spins interact with the V2 electron spin density giving rise to a hyperfine coupled electron-nuclear $S=3/2 \otimes I=1/2$ level system  (Fig. \ref{fig1}c) and we utilise an electron qubit encoding as $|0\rangle = |\pm 1/2\rangle, |1\rangle = |\pm 3/2\rangle$. 

For many of the nearby strongly coupled spins the isotropic Fermi contact term results in an almost diagonal hyperfine tensor \cite{hesselmeier2023measuring}. 
In analogy to the well studied NV system \cite{neumann2010single}, at elevated magnetic fields $B_z$, the Hamiltonian of the system aligned with magnetic field reads: 
\begin{equation}
\begin{split}
H_{g(e)}& = D_{g(e)} \left(S_z^2 + \frac{1}{3}S\left( S+1\right)\right) + \gamma B_z S_z \\
&+ A_{\perp}^{g(e)}\left(S_y I_y + S_xI_x\right) + A_{\|}^{g(e)} S_z I_z   + \gamma_n B_z I_z,
\end{split}
\end{equation}
where $A_{\perp,\|}^{g(e)}$ are the diagonal elements of the hyperfine tensor in ground (excited) state, $D_{g(e)}$ is the zero field splitting in the ground (excited) state.
At high magnetic fields this Hamiltonian yields a robust nuclear qubit encoded in the eigenstates of the $I_z$ operator. 
Its lifetime increases quadratically with the magnetic field \cite{neumann2010single} and allows to probe the quantum state multiple times (see Fig. \ref{fig1}b) to overcome the shot noise of the optical readout. 
Hence, the spin states can be probed faster than spin flips occur as marked by the quantum jumps trajectory presented in Fig. \ref{fig1}d.
Here we continuously probe the system and sum the collected photons of $N$ subsequent reading steps to trace the spin quantum state.
The count rate is directly related to the nuclear spin state and enables observation of quantum jumps in real-time. 
Each reading step contains a $C_nNOT_e$ gate realised via two selective microwave $\pi$-pulses $\pm 1/2\to\pm 3/2$ conditional to the same nuclear spin state, and a short electron state readout via A2 laser pulse. This laser pulse also resets the electron state into $m_s$=$\pm1/2$ for the next iteration (grey line in Fig. \ref{fig1}b).

To optimise the readout parameters we first quantify the intrinsic switching and emission rates under the readout process analogous to \cite{zahedian2023readout}. 
In the first step, we exploit the two-point measurement scheme as described above with an additional charge resonance check (CRC) to filter out individual ionisation and off-resonance events (Fig. \ref{fig2}a).
We prepare the initial state by measurements, i.e. using a low count rate in first readout step (R1) (see below) as state indicator.
By choosing the appropriate condition of the $C_nNOT_e$ gate in the first readout we prepare $m_i = |\uparrow\rangle$ or $m_i = |\downarrow\rangle$.
We probe the nuclear spin state with a second readout step (R2) and record a photon count distribution conditional on the state prepared by the first readout $m_i = |\uparrow (\downarrow)\rangle$ for various readout parameters. 
We depict the case of $N_2 = 300, 700$ $t_l = \SI{15}{\mu s}$ in Fig. \ref{fig2}b, while other cases are presented in detail in SM. 
By fitting the distribution with a numerical model \cite{zahedian2023readout} we extract switching ($\gamma$) and emission ($\lambda$) rates depending on the parameters.
For a readout time $t_l = \SI{15}{\mu s}$ and $I = \SI{20}{nW}$ we observe  $\gamma_0  = \SI{8(2)}{Hz}$ and $\gamma_1  = \SI{100(10)}{Hz}$ switching rates. 
We attribute the fast decay rate from the bright state to the increased nuclear spin flipping rates during the optical excitation cycle. The decay rate from the dark state is limited by the infidelity of nuclear state mapping to the electron state and an occasional unwanted excitation cycle, which is supported by a much slower longitudinal spin relaxation in the absence of readout (see SM).
Next, using our experimentally calibrated switching model we estimate the fidelity of the readout (Fig \ref{fig2}c).
Using the maximum likelihood method \cite{zahedian2023readout} we estimate that with a success rate $\eta=100\%$ of the readout an average fidelity for dark and bright states $\mathcal{F}=92 \%$ is achievable with $N_2 = 250$ repetitions and photon threshold  of $n_{th} =3$. 
A better fidelity could be achieved by increasing the threshold for a bright state readout, but has a limited effect on reading a dark state, due to the finite overlap of bright and dark distribution at $n=0$ photon count. 
We see that a fidelity of $99.5 \%$ is achievable when increasing the threshold for reading the bright state, limited by a success rate $\eta = 10 \%$ of the readout. 
For estimating the state preparation fidelity we use the maximum likelihood method and photon count distribution conditioned on the state at the end of the readout as described in \cite{zahedian2023readout}.
We find that conditioning to $n=0$ photons in this case with fidelity $99\%$ prepares a dark state at $N=300$, while preparation of the bright state with similar fidelity is possible only when using less than $\eta=10\%$ success rate of the readout and using higher threshold values.
Thus in our work, we use the dark state preparation, which, combined with CRC heralds a high fidelity of dark nuclear spin state. 
The readout method with probing only one bright state has a limited overall fidelity, since the dark state readout fidelity could not be improved as discussed before by a stronger postselection and sacrificing of the success rate. 
This can be achieved when using alternating addressing of both nuclear spin states, and in the end comparing the two accumulated count rates $n = n_1 - n_2$. 
In this case both nuclear spin states are bright and allow postselection with higher thresholds. 
We use a switching model based on Skellam distribution (see SM) adapted from \cite{zahedian2023readout} to extract the rates and find that in this case the switching rates are half of those for the bright states for both states, resulting from the fact that both states are probed $50\%$ of the overall readout time. 
The numerical analysis shows that one can achieve $\mathcal{F}=98\%$ fidelity for average fidelity of both states with $\eta=5-10 \%$ success rate. This can be further improved by reducing the success rate, showing that there is a trade-off relation. 

\begin{figure}[t]
	\centering
	\includegraphics[width=\columnwidth]{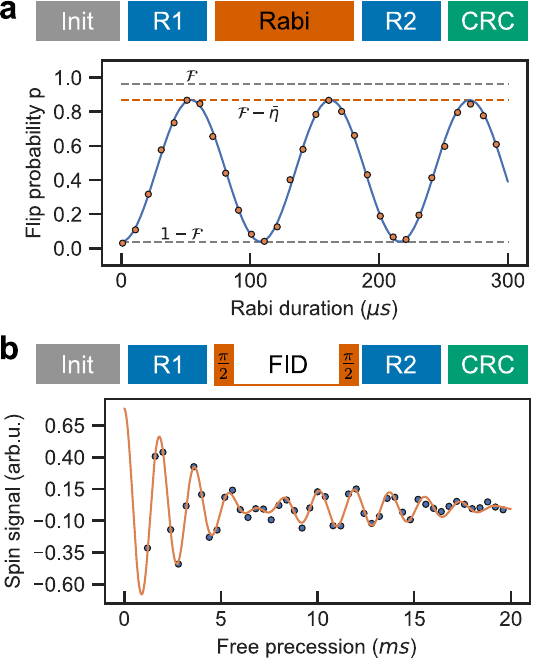}
	\caption{Coherent control of the nuclear spin memory \textbf{(a)} Measurement of the nuclear rabi nutations. (init), Readout 1 (R1), Rabi drive for various duration and Readout 2 (R2), followed by Charge resonance check (CRC),
	\textbf{(b)} Free precession (Ramsey) of the nuclear spin memory}
	\label{fig3}
\end{figure}
We further measure Rabi oscillations of the nuclear spins (Fig. \ref{fig3}a). 
For this we exploit the two point measurement scheme as explained before, and flip the nuclear spin in both states $m_s = \pm 1/2$ with same Rabi frequency. We observe a high nuclear spin contrast, which corresponds to fidelity of $\mathcal{F}=96\%$ for readout with about $\eta=90\%$ success rate.
Further,  we perform nuclear spin Ramsey in electron subdomain $m_s = +1/2$ and obtain $T_2^*=\SI{7.4(4)}{ms}$, a higher value than previously measured at smaller B fields \cite{hesselmeier2023measuring}.
We note an additional beating oscillation, presumably originating to additional coupling to another nuclear spin. 
This could enable in the future to utilise a stronger coupled nuclear spins to address the weaker coupled and obtain nuclear-nuclear $J$-coupling, required for full system Hamiltonian estimation.
 
Finally we apply the efficient readout for probing the weakly coupled nuclear spins. 
We use an electron-nuclear double resonance (ENDOR) pulse sequence \cite{aslam2017nanoscale} for probing the $A_{zz}$ coupled nuclear spins where the nuclear spin is used as memory and a readout ancillary spin and the electron spin as a sensor. 
The sequence exploits the electron-nuclear double resonance condition, and results in a readout contrast increase, once the target spins are flipped. 
We scan the frequency of the probing pulse and observe two distinct location of peaks centred at the positions of the expected Larmour frequency of corresponding nuclear spin baths of $^{29}$Si and $^{13}$C.
We observe 8 individual peaks around $^{29}$Si bath and 4 individual peaks around $^{13}$C bath with copling strength within 10-100 kHz range, suitable for utilising as building blocks of a quantum register. 
\begin{figure}[t]
	\centering
	\includegraphics[width=\columnwidth]{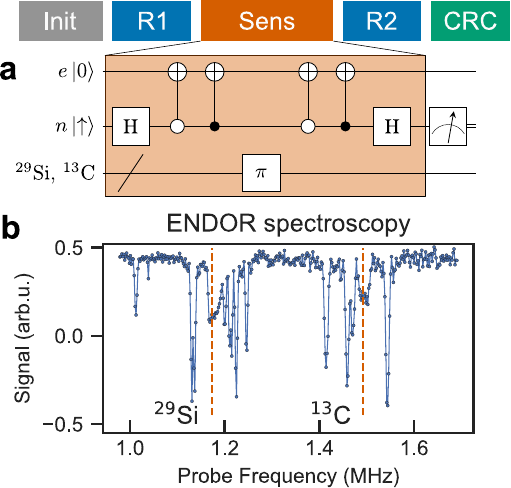}
	\caption{ENDOR spectroscopy of nuclear spin bath \textbf{(a)} Measurement protocol, initialisation and readout are performed similarly to as Fig. \ref{fig3}. The coherent control is performed via entangling with electron spin, acquiring a phase, conditioned on whether the target nuclear spins are probed with a resonant $\pi$ pulse. 
	\textbf{(b)} ENDOR spectra of nuclear spin bath containing $^{13}$C and $^{29}$Si spins. Vertical dashed lines mark the expected Larmor frequency based on the estimated B field.}
	\label{fig4}
\end{figure}

\section{Conclusion}

In this work we show high fidelity readout of a single nuclear spin around V2 center with fidelity reaching 99\%. We demonstrate that such readout can serve for initialisation by measurement with fidelity reaching 99\%. 
We show coherent control of the nuclear spin after initialisation by measurement and demonstrate the application of the high fidelity readout for the efficient sensing of weakly coupled nuclear spins. 
A further improvement in fidelity could be reached by further increasing the B field values, as the lifetime of the spin increases quadratic with B field \cite{neumann2010single}. 
Furthermore, a photonic integration with nanostructures, will allow to increase the number of photons per iteration and allow for high fidelity at smaller B fields. 
Weaker coupled nuclear spins \cite{nizovtsev2018non} will favour higher fidelities at smaller magnetic fields as well. 
Our results mark the important step forward for the qubit candidate system V2 center in 4H-SiC i.e. single shot readout, important for quantum information applications such as quantum information processing, state preparation by measurement and quantum sensing. 



\acknowledgments
P.K., F.K., J.U.H., and J.W. acknowledge support from the European Commission through the QuantERA project InQuRe (Grant agreements No. 731473, and 101017733).
P.K., F.K. and J.W. acknowledge the German ministry of education and research for the project InQuRe (BMBF, Grant agreement No. 16KIS1639K).
F.K. and J.W. acknowledge support from the European Commission for the Quantum Technology Flagship project QIA (Grant agreements No. 101080128, and 101102140), the German ministry of education and research for the project QR.X (BMBF, Grant agreement No. 16KISQ013) and Baden-Württemberg Stiftung for the project SPOC (Grant agreement No. QT-6).
J.W. also acknowledges support for the project Spinning (BMBF, Grant agreement No. 13N16219), and DFG via INST (41/1109-1 FUGG).
J.U.H. further acknowledges support from the Swedish Research Council under VR Grant No. 2020-05444 and Knut and Alice Wallenberg Foundation (Grant No. KAW 2018.0071).

\appendix
\section{Experimental setup}
All experiments were performed at cryogenic temperature $<$\SI{10}{\kelvin} in a Montana Instruments cryostation. 
A self-build confocal microscope was used to optically excite single V2 centers and detect the red-shifted phonon side band. 
Off-resonant excitation was performed with a 728 nm diode laser (Toptica iBeam Smart). 
For resonant excitation we used an external cavity tunable diode laser (Toptica DL Pro), which was split and frequency shifted by two AOMs to address both optical transitions. 
Laser photons are filtered by two tunable long-pass filters (Semrock TLP01-995).
The magnetic field is created via a electromagnet from Danfysik. In this work a magnetic field of $B\sim0.14$ T was used.
For most of our measurements, we used 20nW A2 excitation power before the cryostation. 
The used detectors are fiber coupled superconducting nanowire single-photon detectors from Photon Spot.

\section{Silicon Carbide sample}
The used substrate has natural abundance of silicon (4.7\% $^{29}$Si) and carbon (1.1\% $^{13}$C) isotopes, which are spin $I = 1/2$ nuclei.
\begin{table}[b]
\begin{center}
	\begin{tabular}{|c|c|c|c|c|c|c|}
	\hline 
	Set&$A_{zz}$  &$A_{xx}$ &$A_{yy}$ &$A_{iso}$ & $T$ \\ [0.5ex] 
	\hline
     Si$_{II}$ &8.660(3) &9.00(1) &9.03(1) & 8.910(6) & -0.130(4) \\ [1ex]  
	\hline
	\end{tabular}
	\caption{\label{tab1} The hyperfine tensor of the used nuclear spin. Units are in MHz. T stands for the dipolar interaction term.}
\end{center}
\end{table}
 
%



\section*{Supplementary material}

# Supplemental Material for "High fidelity optical readout of a nuclear spin qubit in Silicon Carbide"

Erik Hesselmeier,[1] Oliver von Berg,[1] Pierre Kuna,[1] Wolfgang Knolle,[2] Florian Kaiser,[3, 4] Nguyen Tien Son,[5] Misagh Ghezellou,[5] Jawad Ul-Hassan,[5] Vadim Vorobyov,[1, *] and Jörg Wrachtrup[1, 6]

[1] *3rd Institute of Physics, IQST, and Research Centre SCoPE, University of Stuttgart, Stuttgart, Germany*
[2] *Department of Sensoric Surfaces and Functional Interfaces,*
*Leibniz-Institute of Surface Engineering (IOM), Leipzig, Germany*
[3] *Materials Research and Technology (MRT) Department,*
*Luxembourg Institute of Science and Technology (LIST), 4422 Belvaux, Luxembourg*
[4] *University of Luxembourg, 41 rue du Brill, L-4422 Belvaux, Luxembourg*
[5] *Department of Physics, Chemistry and Biology, Linköping University, Linköping, Sweden*
[6] *Max Planck Institute for solid state physics, Stuttgart, Germany*

## I. RATE MODEL FOR THE SINGLE SIDED READOUT

We fit the photon count distribution with a switching model of two Poisson distributed with average $\lambda_1$ and $\lambda_2$ [1] for estimating the switching and emission rates of the readout:

$$p\left(\lambda || \uparrow\right) = \int_0^T \left\{ e^{(\gamma_1 - \gamma_2)(T-\tau) - \gamma_1 T} \left( \gamma_2 I_0 \left( 2\sqrt{\gamma_1 \gamma_2 \tau (T-\tau)} \right) + \sqrt{\frac{\gamma_1 \gamma_2 (T-\tau)}{\tau}} I_1 \left( 2\sqrt{\gamma_1 \gamma_2 \tau (T-\tau)} \right) \right) \times \right.$$
$$\left. \text{Poisson} \left( \lambda, \lambda_1 \tau + \lambda_2 (T-\tau) \right) d\tau \right\} + e^{-\gamma_2 T} \text{Poisson} \left( \lambda, \lambda_2 T \right) \tag{1}$$

$$p\left(\lambda || \downarrow\right) = \int_0^T \left\{ e^{(\gamma_2 - \gamma_1)\tau - \gamma_2 T} \left( \gamma_1 I_0 \left( 2\sqrt{\gamma_1 \gamma_2 (T-\tau)} \right) + \sqrt{\frac{\gamma_1 \gamma_2 \tau}{T-\tau}} I_1 \left( 2\sqrt{\gamma_1 \gamma_2 (T-\tau)} \right) \right) \times \right.$$
$$\left. \text{Poisson} \left( \lambda, \lambda_1 \tau + \lambda_2 (T-\tau) \right) d\tau \right\} + e^{-\gamma_1 T} \text{Poisson} \left( \lambda, \lambda_1 T \right) \tag{2}$$

## II. RATE MODEL FOR THE DOUBLE SIDED READOUT

The analytical expression for fitting of the alternating addressing scheme is derived as following. Similar to the single state addressing we follow a switching telegraph process [1]. We replace the integration of emitting a $\mu = \lambda_0 \tau + \lambda_1 (T-\tau)$ Poisson distributed photons with a Skellam distribution with $\mu_1 = \lambda_0 \tau + \lambda_1 (T-\tau)$ and $\mu_2 = \lambda_1 \tau + \lambda_0 (T-\tau)$.

$$p\left(\lambda || \uparrow\right) = \int_0^T \left\{ e^{(\gamma_1 - \gamma_2)(T-\tau) - \gamma_1 T} \left( \gamma_2 I_0 \left( 2\sqrt{\gamma_1 \gamma_2 \tau (T-\tau)} \right) + \sqrt{\frac{\gamma_1 \gamma_2 (T-\tau)}{\tau}} I_1 \left( 2\sqrt{\gamma_1 \gamma_2 \tau (T-\tau)} \right) \right) \times \right.$$
$$\left. \text{Skellam} \left( \lambda, \lambda_1 \tau + \lambda_2 (T-\tau), \lambda_2 \tau + \lambda_1 (T-\tau) \right) d\tau \right\} + e^{-\gamma_2 T} \text{Skellam} \left( \lambda, \lambda_2 T, \lambda_1 T \right) \tag{3}$$

$$p\left(\lambda || \downarrow\right) = \int_0^T \left\{ e^{(\gamma_2 - \gamma_1)\tau - \gamma_2 T} \left( \gamma_1 I_0 \left( 2\sqrt{\gamma_1 \gamma_2 (T-\tau)} \right) + \sqrt{\frac{\gamma_1 \gamma_2 \tau}{T-\tau}} I_1 \left( 2\sqrt{\gamma_1 \gamma_2 (T-\tau)} \right) \right) \times \right.$$
$$\left. \text{Skellam} \left( \lambda, \lambda_1 \tau + \lambda_2 (T-\tau), \lambda_2 \tau + \lambda_1 (T-\tau) \right) d\tau \right\} + e^{-\gamma_1 T} \text{Skellam} \left( \lambda, \lambda_1 T, \lambda_2 T \right) \tag{4}$$

---

* v.vorobyov@pi3.uni-stuttgart.de

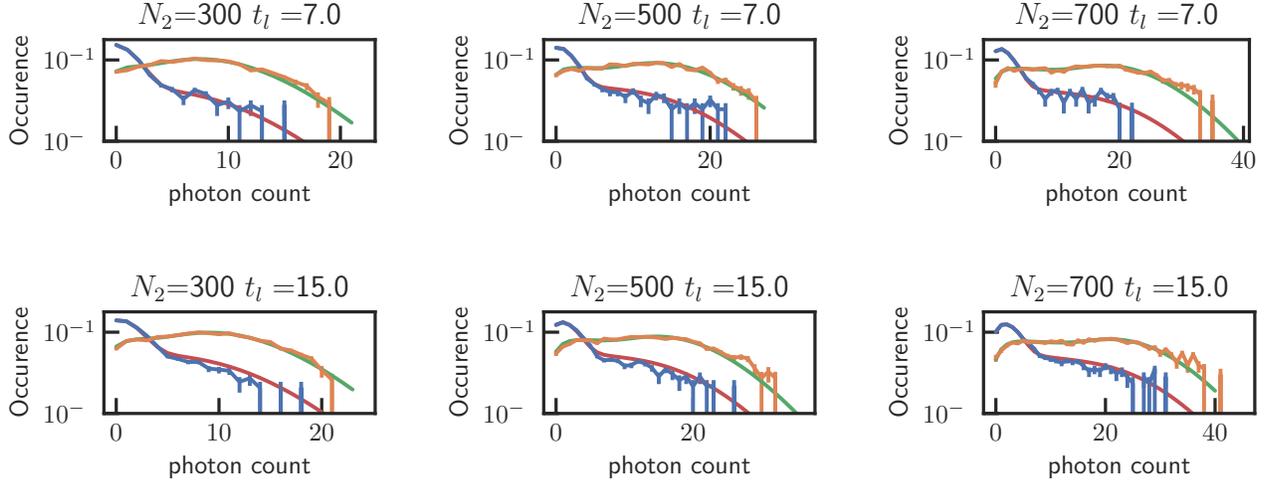

FIG. 1. Distribution of the photon count based on prepared $|\uparrow\rangle$ and $|\downarrow\rangle$ states

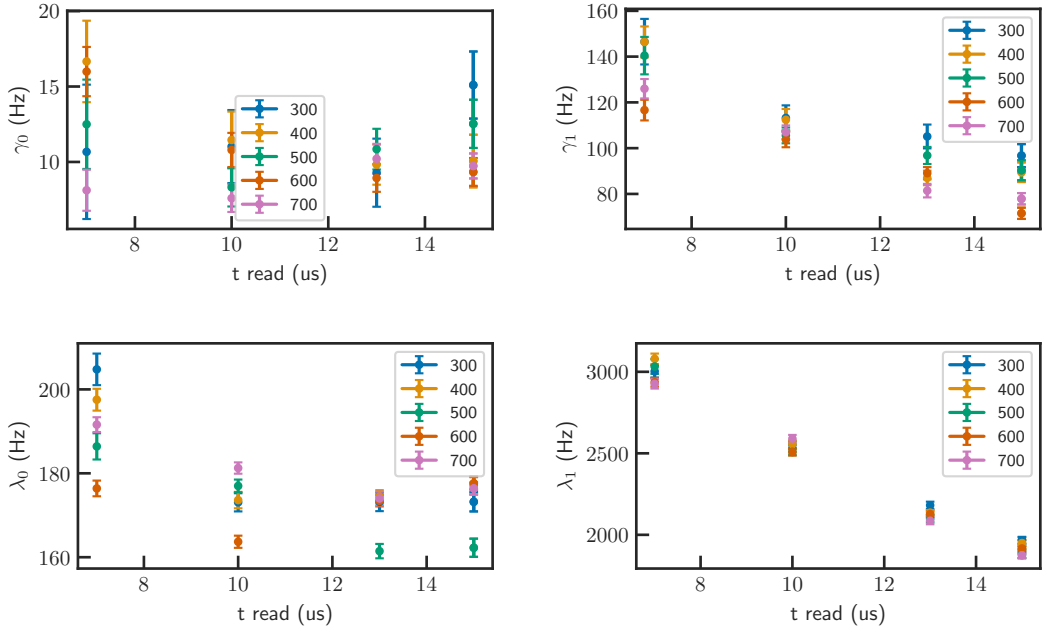

FIG. 2. Switching ($\gamma$) and emission ($\lambda$) rates of the nuclear spin state under the readout technique presented in main text fig2a

As Skellam distrubtion defines a random variable, which is equal to a difference of two Poisson distributed variables and naturally follows from definition of our double sided readout. The fitted distributions

## III.   DECAY OF THE MEMORY

We probe the T1 time of the memory in the absense of the readout. We perform a measurement up to 1s and see to visible decay.

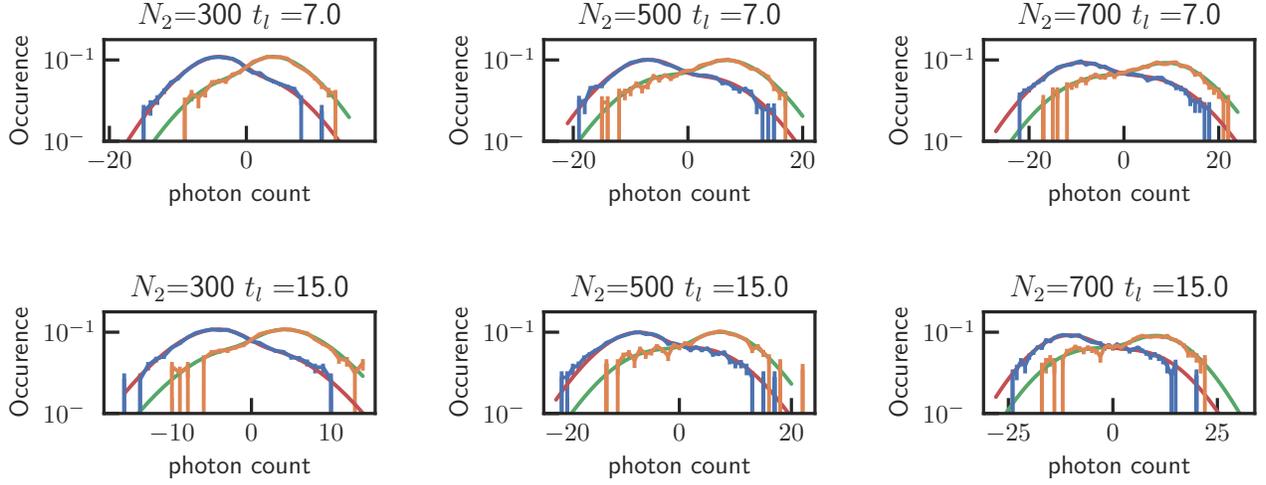

FIG. 3. Distribution of the photon count in alternating addresing scheme. The photon count is a difference of addressing $|\uparrow\rangle$ and $|\downarrow\rangle$ states

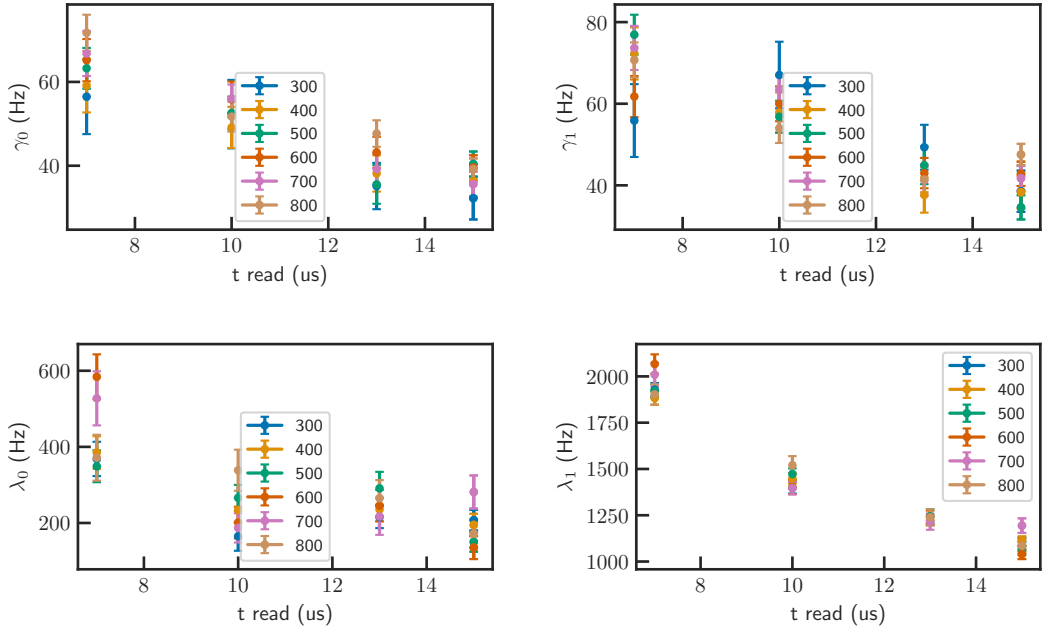

FIG. 4. Switching ($\gamma$) and emission ($\lambda$) rates of the nuclear spin state under the alternating addressing readout technique presented in main text fig2e




\end{document}